\documentclass[usenatbib]{mn2e}
\usepackage{epsfig}
\usepackage{amsmath}
\usepackage{ulem}

\def\gsim{\mathrel{\raise0.35ex\hbox{$\scriptstyle >$}\kern-0.6em 
\lower0.40ex\hbox{{$\scriptstyle \sim$}}}}
\def\lsim{\mathrel{\raise0.35ex\hbox{$\scriptstyle <$}\kern-0.6em 
\lower0.40ex\hbox{{$\scriptstyle \sim$}}}}

\def\msun{{\rm M}$_{\odot}$}

\date{\today}
\title[Faint End of CMR at z$\sim 0.8$ cluster]
{Dependence of the Build-up of the Colour-Magnitude Relation on Cluster
Richness at $z \sim 0.8$}

\author[Y. Koyama et al.]{
\parbox[t]{\textwidth}{
Yusei Koyama$^{1}$\thanks{E-mail: koyama@astron.s.u-tokyo.ac.jp},
Tadayuki Kodama$^{2}$,
Masayuki Tanaka$^{1,4}$,
Kazuhiro Shimasaku$^{1,3}$ and
Sadanori Okamura$^{1,3}$
}
\vspace*{6pt}\\
$^{1}$Department of Astronomy, School of Science, The University of Tokyo,
Tokyo 113-0033, Japan\\
$^{2}$National Astronomical Observatory of Japan, Mitaka, Tokyo 181-8588, Japan\\
$^3$Research Center for Early Universe, School of Science, The
University of Tokyo, Tokyo 113-0033, Japan \\
$^4$European Southern Observatory, Karl-Schwarzschild-Str. 2, D-85748, Garching
bei M\"{u}nchen, Germany
}
\begin{document}

\maketitle

\begin{abstract}

We present environmental dependence of the build-up of the
colour-magnitude relation (CMR) at $z \sim 0.8$. It is well 
established that massive early-type galaxies exhibit a tight 
CMR in clusters up to at least $z \sim 1$.
The faint end of the relation, however, has been much less explored 
especially at high redshifts primarily due to limited depths of the data.
Some recent papers have reported a deficit of the faint red galaxies 
on the CMR at $0.8 \lsim z \lsim 1$, but this has not been well confirmed yet 
and is still controversial.
Using a deep, multi-colour, panoramic imaging data set of the distant 
cluster RXJ1716.4+6708 at $z=0.81$, newly taken with the Prime Focus 
Camera (Suprime-Cam) on the Subaru Telescope, we carry out an analysis 
of faint red galaxies with a care for incompleteness.  
We find that there is 
a sharp decline in the number of red galaxies toward the faint end of
the CMR below $M^*+2$.
We compare our result with those for other clusters at $z \sim 0.8$ 
taken from the literature, which show or do not 
show the deficit. 
We suggest that the "deficit" of faint red galaxies is dependent 
on the richness or mass of the clusters, in the sense that poorer 
systems show stronger deficits.  This indicates that the evolutionary 
stage of less massive galaxies depends critically on environment.

\end{abstract}
\begin{keywords}
galaxies: clusters: individual: RXJ1716.4+6708 ---
galaxies: evolution ---
galaxies: luminosity function, mass function

\end{keywords}
\section{Introduction}
\label{sec:intro}

It is well known that red early-type galaxies exhibit a tight 
sequence on colour-magnitude diagrams, which is called the
colour-magnitude relation (CMR) (e.g., Visvanathan \& Sandage 1977;
\citealp{bow92}). In nearby clusters, the CMRs extend down to at least $5-6$
magnitude fainter than the brightest cluster galaxies 
(e.g., Terlevich, Caldwell \& Bower 2001).
The small colour scatter around the CMR is indicative of the homogeneity
of early-type galaxies in clusters (e.g., Bower et al.\ 1992, 1998).
At high redshifts, the CMR has already been well established in clusters
at least out to $z\sim1$ as far as the bright end is concerned
(e.g., \citealt{ell97}; \citealt{kod98}; \citealt{sta98};
\citealp{van98}; 2001; Blakeslee et al.\ 2003; Stanford et al.\ 2006;
Mei et al.\ 2006a,b). The faint end of the CMR, however, 
has been much less explored and still highly uncertain.

Some recent deep studies of distant galaxy clusters have shown 
a relatively small number of galaxies at the faint end of the CMR
compared to local clusters. De Lucia et al.\ (2004, 2007) showed 
such a deficit of faint red galaxies in $z=0.6-0.8$ clusters 
observed by the ESO Distant Cluster Survey (EDisCS; \citealt{whi05}).
A similar result was shown in Stott et al.\ (2007). They compared the 
faint end of the luminosity functions of red galaxies in $z\sim 0.5$
clusters from Massive Cluster Survey (MACS; Ebeling et al. 2001) with
those of $z \sim 0.1$ clusters from Las Campanas/AAT Rich Cluster Survey
(LARCS; Pimbblet et al. 2006).   
\cite{tan05} analysed the RXJ0152.7--1357 cluster (hereafter RXJ0152)
at $z=0.83$ based on wide field data taken with the Subaru Prime Focus 
Camera on the Subaru Telescope (Suprime-Cam; Miyazaki et al.\ 2002), 
and they also showed a deficit of faint red galaxies on the CMR.
Based on these results, De Lucia et al. (2004, 2007) and \cite{tan05} 
discussed that the faint end of the CMR well visible in the 
present-day universe was established at relatively later epochs as 
faint blue galaxies stopped their star formation after $z\sim0.8$ 
in contrast to much earlier ($z\gg1$) termination of star formation 
in massive galaxies. 
\cite{tan05} classified galaxy environment 
into ``cluster'', ``group'' and ``field'', and examined the 
environmental dependence of the faint end of the CMR as well. 
They suggest that the build-up of the CMR depends also on 
environment in the sense that it is more delayed in lower-density
environment.

The deficit of the faint end of the CMR is often discussed 
in the line of a currently favoured observational phenomenon 
called ``down-sizing''. This trend was first noted for field galaxies 
by Cowie et al.\ (1996). They showed in their Hawaii Deep Field 
that most massive galaxies tend to show low star formation rates 
while less massive galaxies still show on-going star formation
activity at $z\lsim1$. 
Such a trend has been extended in both redshift space and magnitude range.
Kauffmann et al.\ (2003) showed in the local SDSS data that massive galaxies
are dominated by red old galaxies. By contrast, less massive galaxies 
show bluer colours due to some on-going star formation, and galaxies 
below a few times $10^{10}$ \msun {} in stellar mass are predominantly blue.
A very similar trend was reported at $z\sim1$ by Kodama et al.\ (2004). 
They looked in the Subaru/XMM Deep Field and showed the distribution 
of galaxies at $z\sim1$ on the colour-magnitude diagram. 
A clear bi-modality on the colour-magnitude diagram
was observed again. Since then, a large number of papers have discussed this
interesting issue. One of the most convincing cases is based 
on $\sim$8,000 galaxies with spectroscopic
redshifts within $0.7<z<1.4$ in the DEEP2 survey (Bundy et al.\ 2006). 
They derived stellar mass functions of red and blue galaxies and showed 
that the mass where the dominant contribution is switched from red 
to blue galaxies shifts to smaller masses as time goes on.
This down-sizing trend is found also in clusters as already mentioned above.

Recently, however, Andreon (2006) claimed that the faint end of the CMR 
is fully in place in the rich cluster MS1054--0321 (hereafter MS1054)
at $z=0.83$ and questioned the universality of the deficiency of 
faint red galaxies at $z \sim 0.8$. A critical problem is that the 
number of galaxy clusters having deep enough imaging data so that 
we can discuss the faint end of the CMR is still very limited at 
high redshifts. In fact, so far only a few clusters at $z \sim 0.8$ 
(of which some are optically-selected clusters) 
have been studied in this respect (i.e., MS1054 by \citealt{and06},
RXJ0152 by \citealt{tan05}, and some optically-selected clusters by
EDisCS in \citealt{del04}, 2007).
Therefore, it is crucial to increase the number of clusters 
and see if the deficit of faint red galaxies is universally 
observed or not and see what determines the degree of the deficit.
In this paper, we examine another cluster at $z=0.81$, 
RXJ1716.4+6708 (hereafter RXJ1716), in order to obtain a more 
general picture of $z\sim0.8$ clusters. We will also discuss a 
possible origin of the cluster-to-cluster variations.

The structure of this paper is the following. In Section 2, we introduce 
our PISCES project, and also summarize the properties of the RXJ1716
cluster shown by some previous works. 
We present a deficit of faint red galaxies in Section 3, and we discuss
the environmental dependence of the nature of faint galaxies in Section 4.
Finally, we summarize our results in Section 5.
Throughout this paper we use $\Omega_M =0.3$, $\Omega_{\Lambda} =0.7$, and 
$H_0 =70$ km s$^{-1}$Mpc$^{-1}$.
Magnitudes are all given in the AB system, unless otherwise stated.

\section{Data}
\label{sec:data}

\subsection{PISCES project}
\label{sec:pisces}

Our new data for the RXJ1716 cluster which we use in this paper
were obtained as part of the PISCES programme (Panoramic Imaging and
Spectroscopy of Cluster Evolution with Subaru) and  
the details of this project are given in Kodama et al.\ (2005). 
Here we repeat only the basic concepts of the project.
The Subaru Telescope has both great light-collection power and 
superb image quality.  At the same time, 
the Subaru Prime Focus Camera (Suprime-Cam; Miyazaki et al.\ 2002)  
has a very wide field of view of $34' \times 27'$.  
This unique combination has enabled us to conduct a deep and wide study of
distant galaxy clusters out to $z \sim 1.3$. In particular, we are able to
view from cluster cores to the surrounding general field all at once.
Our aim is to probe when and where cluster galaxies form, and how 
galaxies evolve afterwards depending on environment and galaxy mass.
We study more than 10 X-ray detected distant clusters at various
redshifts ($0.4 \lsim z \lsim 1.3$), hence at various stages, of galaxy
evolution (Kodama et al.\ 2005).
We first map out large scale structure on a comoving scale of 10--15~Mpc
based on photometric redshifts, and then by studying the properties of
galaxies as functions of environment and time in detail, 
we try to make a link between the evolution of galaxies and the growth of
large scale structures through environmental effects.
Some clusters have already been observed and analysed and the results
have been published in several papers
(e.g., Kodama et al.\ 2001, 2004, 2005; Tanaka et al.\ 2005, 2006, 2007ab;
Nakata et al.\ 2006).
 
\subsection{Observation and Data reduction}
\label{sec:obs}
We obtained an imaging data set of RXJ1716 
on 2005 May 5 and 6. Properties of this cluster are summarized in Section 2.3. 
The observing conditions were very good, and the seeing size was stable
at $\sim 0''.7$ during the nights.  We observed this cluster
in the $VRi'z'$ filters.
Exposure times and limiting magnitudes are shown in Table~1.
\begin{table}
\begin{center}
\begin{tabular}{c|c|c}
\hline
\hline
Filter  & Exposure time (min)  & Limiting magnitude \\ \hline
$V$  & 85    & 26.6   \\ 
$R$  & 85    & 26.3   \\
$i'$ & 46    & 25.6   \\
$z'$ & 51    & 24.9   \\
\hline
\end{tabular}
\caption{Exposure times and limiting magnitudes (AB) of the Suprime-Cam
data.  Limiting magnitudes correspond to 5$\sigma$ detection limits
and measured in a 2$''$ aperture.}
\label{tab:data}
\end{center}
\end{table}
The data were reduced with the {\sc sdfred} software
(\citealt{yag02}; \citealt{ouc04}). 
Catalogues were created using the {\sc sextractor} software \citep{ber96}.
Objects were selected at $z'$-band.
We used {\sc mag\_auto} for the total magnitudes of the objects, and {\sc
mag\_aper}
measured in a 2$''$ aperture (corresponding to 15~kpc at the cluster redshift)
for the colours of the objects. 
 \begin{figure*}

   \begin{center}
    \leavevmode
    \rotatebox{0}{\includegraphics[width=11cm,height=11cm]{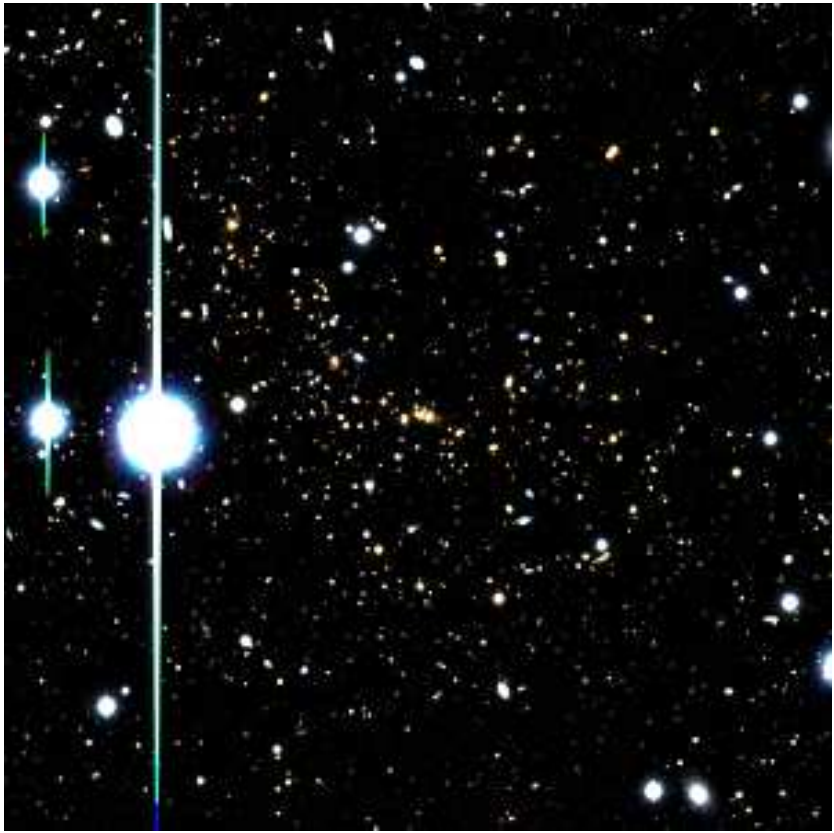}}
   \end{center} 
   \caption{False-colour image of the central $5'\times 5'$ region
of the RXJ1716 cluster constructed from our $R$, $i'$ and $z'$
images.  North is up and east is to the left. 
 \label{fig:image}}
 \end{figure*}
\subsection{RXJ1716.4+6708 cluster (RXJ1716)}
The cluster was first discovered in the ROSAT North Ecliptic Pole Survey
(Henry et al.\ 1997).
Gioia et al.\ (1999) performed optical spectroscopy and
identified 37 cluster members, from which the central redshift
of RXJ1716 is measured to be 0.81. 
It has been known that this cluster has a small 
subcluster or group to the northeast of the main cluster, and
the main body of the cluster is elongated toward the subcluster 
(e.g., Jeltema et al.\ 2005). \cite{clo98} noted
that the brightest cluster galaxy (BCG) of this cluster 
is located on the northwestern edge of the structure. 
The galaxy distribution in the optical images resembles
an inverted S-shaped filament (Gioia et al.\ 1999).
The colour image of the central region of RXJ1716 constructed 
from our data is shown in Fig.\ 1.
As claimed in Gioia et al.\ (1999), red galaxies are distributed 
from the cluster core toward the north-east direction. 
This filamentary structure is also seen in the X-ray image shown in 
Jeltema et al.\ (2005).

RXJ1716 has a rest-frame X-ray luminosity of 
$L_{bol} = 13.86 \pm 1.04  \times 10^{44} $erg s$^{-1}$, and 
the temperature is estimated to be $kT = 6.8 ^{+1.0}_{-0.6}$~keV
based on Chandra data (Ettori et al.\ 2004,
see also Gioia et al.\ 1999, Vikhlinin et al.\ 2002, and Tozzi et al.\ 2003
for other measurements of the hot gas temperature).
In this paper, we use the X-ray data from Ettori et al.\ (2004). 
This is because Ettori et al.\ (2004) studied RXJ0152
and MS1054 (both at $z=0.83$) in the same way as RXJ1716, 
so that we can make a fair comparison between these clusters (see Section 4).
Note, however, that a different choice of X-ray luminosity or temperature
would not affect our conclusions.

In Gioia et al.\ (1999), the velocity dispersion of this cluster 
is estimated to be $1522 ^{+215} _{-150}$ km s$^{-1}$  using their
spectroscopic data, and they noted that the velocity dispersion 
of this cluster is higher than expected from its temperature. 
They therefore suggest that RXJ1716 may not have reached  
a virial equilibrium. 
The weak-lensing mass is estimated to be 
$2.6 \pm 0.9 \times 10^{14} h^{-1} M_{\odot}$ (\citealt{clo98}). 
This is consistent with the mass estimation from
the X-ray data in \cite{ett04}, 
$M_{\rm{tot}} = 4.35 \pm 0.83  \times 10 ^{14} M_{\odot}$. 

\section{Results}
\label{sec:results}

Our results are divided into three parts.  First of all, we map out the 
structures in and around the cluster by tracing the member candidates
selected on the basis of photometric redshifts.
Secondly, we show colour-magnitude diagrams of the galaxies in the
cluster region.  Finally, we draw the luminosity function
of red galaxies in the cluster, and focus on the ``deficit'' of
faint red galaxies by quantifying a luminous-to-faint ratio of the
red sequence galaxies.

\subsection{Large-Scale Structure}
\label{subsec:largescale}
Our first step is to identify a large scale structure of the cluster
from the 2-D distribution of member galaxy candidates.
For this purpose, we apply a photometric redshift technique based on
our $VRi'z'$ photometric data set and the photometric redshift code
by Kodama, Bell \& Bower (1999), and largely eliminate fore-/background 
galaxy contaminations while keeping most of the true cluster members.
In Fig.~2, photometric redshifts are plotted against the
spectroscopic redshifts for 33 
spectroscopically confirmed cluster members given in
Gioia et al.\ (1999).
We can see that most of the known cluster members ($\sim$ 95\%) are assigned
photometric redshifts within the range of $0.76 \le z_{phot} \le 0.83$.
The distribution of photometric redshifts has an asymmetric shape
which has a tail to lower redshifts, due to the inherent problem of
our photometric redshifts (see Kodama et al.\ 1999).
 \begin{figure}
   \begin{center}
    \leavevmode
    \rotatebox{0}{\includegraphics[width=7.0cm,height=7.0cm]{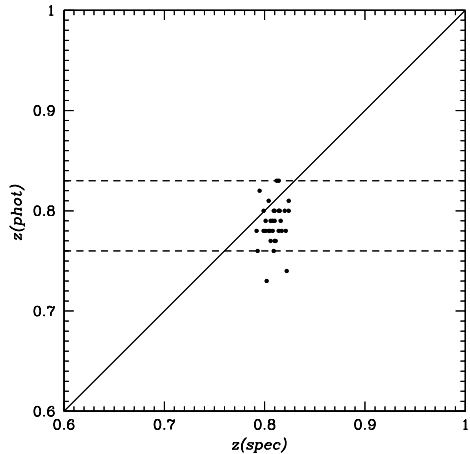}}
   \end{center} 
   \caption{Our photometric redshifts plotted against
the spectroscopic redshifts for the spectroscopically confirmed
cluster member galaxies given in Gioia et al.\ (1999). 
The two horizontal dashed lines indicate the phot-z selection criteria 
to trace the large-scale structure (see text). }
   \label{fig:photz}
 \end{figure} 
Taking this asymmetric distribution into account,
we hereafter consider this redshift range (as shown by two horizontal 
dashed lines) as an appropriate photometric redshift range to trace
large scale structures.
After this process, only $\sim 2,500$ galaxies remained out of
$\sim 45,000$ galaxies detected in our images.
Therefore, this technique is indeed very efficient in eliminating many
fore-/background contaminations.
Fig.~3 shows the distribution of galaxies at 
$0.76 \le z_{phot} \le 0.83$
in the entire field, and Fig.~4 is a close-up view of the central
$10' \times 10'$, where density contours are overlaid.
We have confirmed that a similar structure is recovered even if we
use simple colour cuts in $R-z'$ and $i'-z'$ and isolate the red
sequence galaxies, instead of applying the photometric redshift cut.
Therefore, the structures we see in Figs.\ 3 and 4 are robust and are
not an artefact of photometric redshifts.

 \begin{figure*}   
  \begin{center}
    \leavevmode
    \rotatebox{0}{\includegraphics[width=12cm,height=12cm]{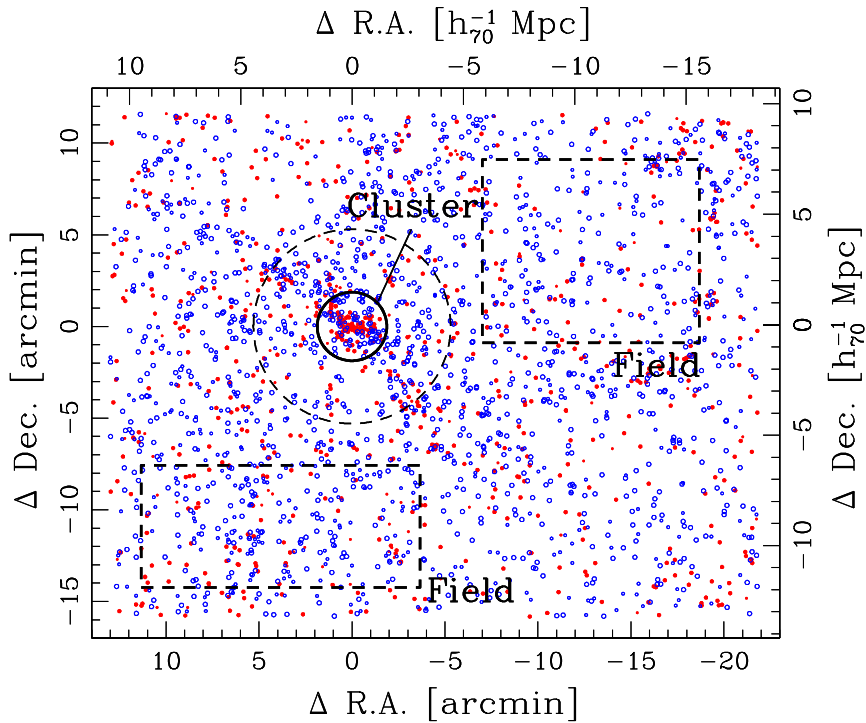}}
   \end{center} 
   \caption{
A 2-D distribution of galaxies around the cluster redshift selected
on the basis of photometric redshifts ($0.76 \le z_{phot} \le 0.83$).
Objects brighter than $m(z') \le 25.0$ are plotted.
The coordinates are shown relative to the centre of the main cluster
$(\alpha =17^{h}16^{m}49^{s}$ and $ \delta =67^{\circ}08'22''$ in J2000).
The top and right panel ticks show the comoving scales in units of Mpc.
The cluster region is defined by the solid circle which corresponds to
$0.35 \times r_{200}$ from the cluster core, while the dashed circle
indicates $r_{200}$. The control field regions are defined by the two dashed
rectangles. Galaxies are plotted as the filled or open
symbols based on their $R-z'$ colour (see text), and their
large and small sizes mean $m(z') < 23.5 $ and $m(z') \ge 23.5$,
respectively. 
   \label{fig:map}}
 \end{figure*} 
 \begin{figure}   
  \begin{center}
    \leavevmode
    \rotatebox{0}{\includegraphics[width=7.0cm,height=7.0cm]{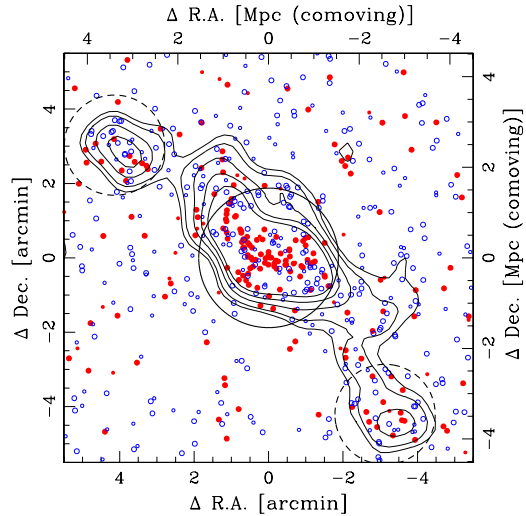}}
   \end{center} 
\caption{A close-up view of the cluster centre ($10' \times 10'$).
The cluster member candidates selected by photometric redshifts are shown.
The contours show the local 2-D number density of
galaxies at 1.5, 2, 3, 4, 5 $\sigma$ above the mean density of the
entire field.
We apply gaussian smoothing (sigma=0.2~Mpc) on each galaxy and combine
the tails of gaussian wings to measure local density at a given point.
A bin size of 0.1~Mpc (physical) is used to draw the iso-density contours.
The coordinates are given with respect to the centre of the main cluster.
The central large circle indicates the definition of the ``cluster''
region (which corresponds to $0.35 \times r_{200}$), and the other two 
circles show the ``group'' regions. 
\label{fig:center_map}}
 \end{figure}

The member candidates plotted in Figs.\ 3 and 4
are divided into red and blue galaxies according to their observed colours.
The red galaxies are defined as $R-z' \ge 2.51-0.049 \times z'$,
and the blue ones are defined as $R-z' < 2.51-0.049 \times z'$.
This boundary is located at 0.2~mag bluer than the best-fitting CMR
(see Section~3.2).
In Fig.~\ref{fig:center_map}, 
we can clearly see two sub-clumps in addition to the elongated cluster core.
One is located to the north-east from the core, and the other is to the 
south-west.
The former one was already suggested in previous works 
(e.g., Gioia et al.\ 1999),
but the latter one has been unknown and this paper reports its first discovery.
Connecting the two sub-clumps and the main body, we can clearly identify a
filamentary structure running across the cluster from north-east to south-west.
The direction is consistent with the elongated structure seen in the 
X-ray images in \cite{jel05}.
By comparing the distributions of red and blue galaxies in Fig.~\ref{fig:map},
we can see that the red galaxies are strongly clustered in the core, 
while the blue galaxies are much less so.
The new group to the south-west and the overall filamentary structure
reported here are based only on photometric redshifts, and they are subject
to projection effects to some extent.
We aim to confirm these structures spectroscopically in our future work.

Based on these structures, we now define galaxy environments,
``cluster (core)'', ``group'' and ``field''.
The characteristic radius $r_{200}$ is defined as the radius 
within which the mean matter density is 200 times larger than
the mean density of the Universe,
and can be calculated as:
$$r_{200} = \frac{\sqrt{3} \sigma}{10 H(z)} $$
(\citealt{car97}).
For RXJ1716, using its velocity dispersion $\sigma=1522$ km s$^{-1}$
from \cite{gio99}, $r_{200}$ is estimated to be $5.3$~arcmin or $2.4$~Mpc 
in physical scale (shown by the dashed circle in Fig.~\ref{fig:map}).  
However, this radius is too large for the apparent extent of the 
cluster galaxies shown in Figs.~3 and 4.
The subclumps and general field are largely included in this circle.
In fact, since this cluster has a relatively large velocity
dispersion for its X-ray luminosity and hence it may not have been
viriarized (\citealt{gio99}; see Section 2.3), the above estimate of
$r_{200}$ may be an overestimate.
Therefore, we define the ``cluster'' region shown by the solid 
circle in Fig.~3, which corresponds to $0.35 \times r_{200}$ ($\sim 0.8
$ Mpc in physical scale).
We should note that our conclusions do not strongly change
if we change the definition of cluster radius within $(0.35 - 1.0)
\times r_{200}$, although the statistical errors become larger when
we take longer radius.

``Group'' regions are taken from the two sub-clumps already mentioned above.
The central coordinates of the groups are (4$'$.14, 3$'$.03) and
($-$3$'$.13, $-$4$'$.21) and a radius of 1.35~arcmin ($\sim 0.6$ Mpc 
in physical scale) is adopted in both groups as shown in Fig.\ 4.

The ``field'' regions are chosen somewhat arbitrarily but avoiding the
filamentary structures and are shown by two rectangles in Fig.~\ref{fig:map}.
These fields will be used as control fields to estimate the contribution of
contaminant galaxies in the cluster and the group regions.

\subsection{Colour-Magnitude Diagrams}
\label{subsec:CMD}

Before we proceed to an analysis of the colour-magnitude diagrams 
and the luminosity function, 
we have to stress that we do not apply the photometric redshift 
selection in the following sections. 
This is because the photometric redshifts of faint galaxies 
are less accurate due to relatively large photometric errors.
In particular, estimation of photometric redshift
is very difficult for some faint galaxies which are not detected 
in some of the $VRi'$ bands. 
Although the fraction of these missed faint galaxies is not large,
we try not to underestimate the number of faint member galaxies as
much as possible for secure conclusions.

\begin{figure*}
\begin{center}
\leavevmode
\begin{center}
  \leavevmode
  \epsfxsize 0.48\hsize
  \epsfbox{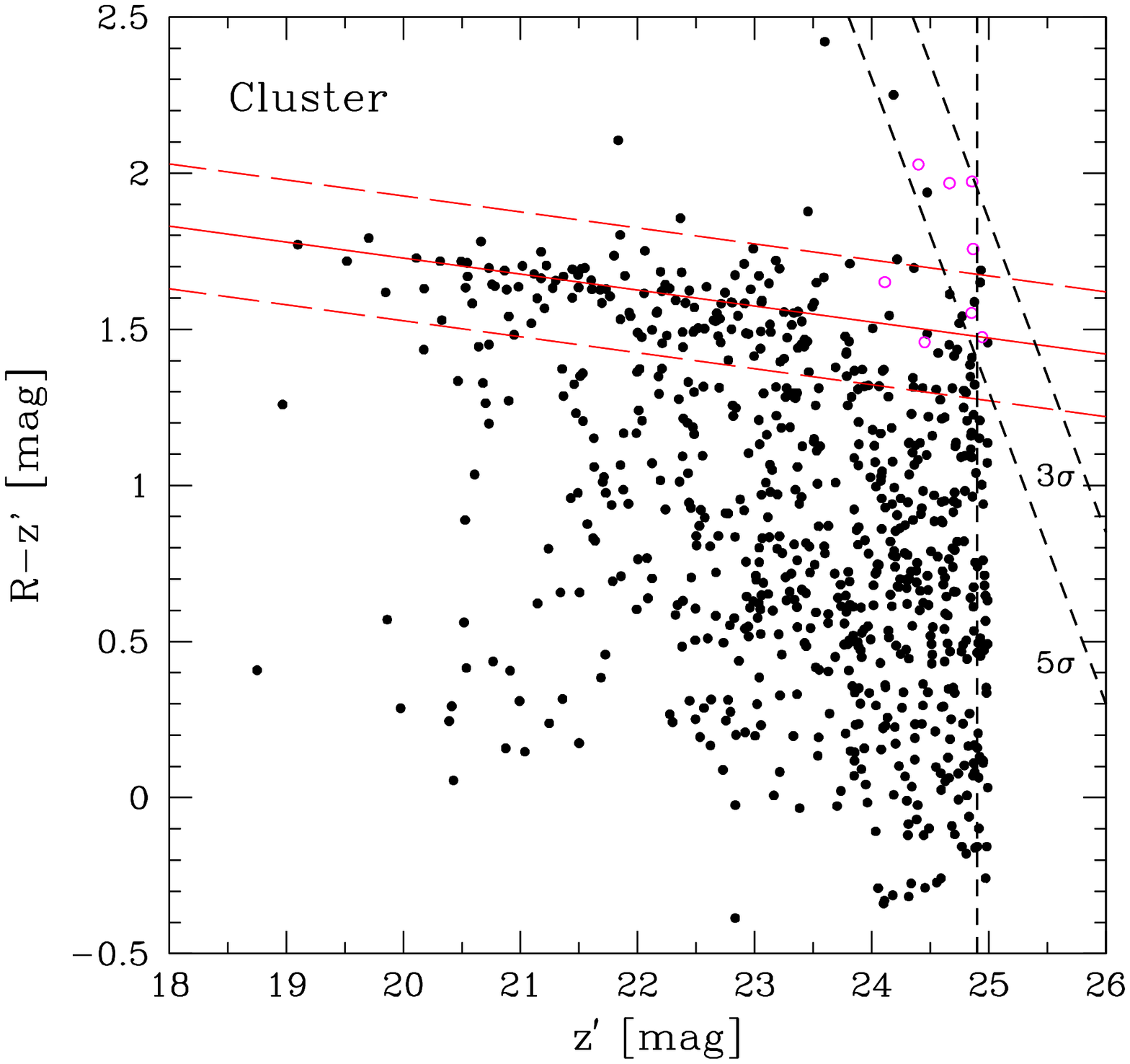}
  \epsfxsize 0.48\hsize
  \epsfbox{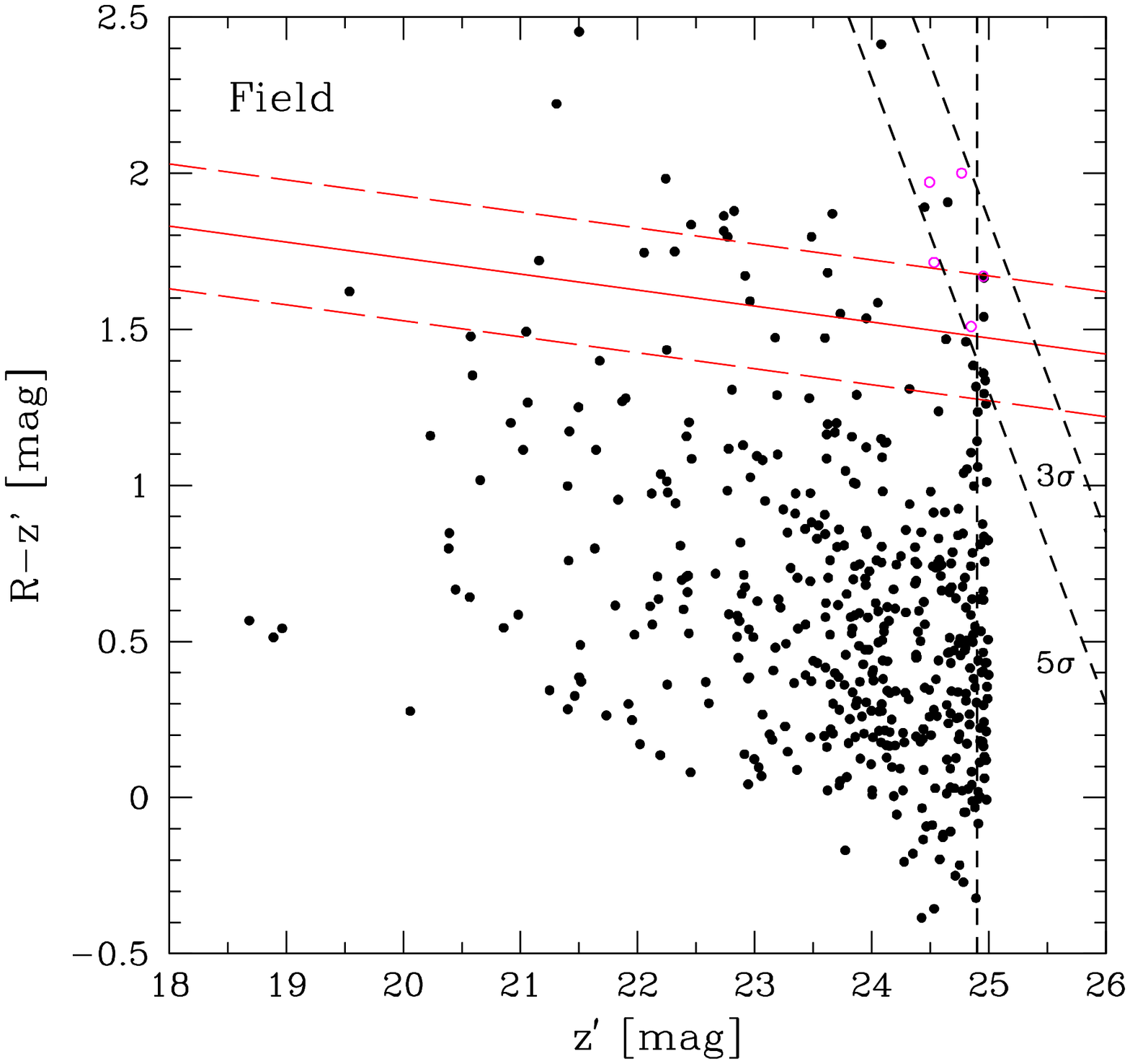}
\end{center}
\end{center}
\caption{Colour-magnitude diagrams ($R-z'$ vs. $z'$)
in the cluster region ({\it left}) and 
the control field region ({\it right}). Objects brighter than
$z'=25.0$ are plotted.
In the right panel, field galaxies are randomly sampled
from the control fields to scale down the surface area and properly
match it to the area of the cluster region.
Photometric redshift selection is not applied.
A clear CMR is seen in the cluster region. The vertical dashed line
shows the 5$\sigma$ detection limit,
and the slanted dashed lines are the $3\sigma$ and $5\sigma$ colour 
limits as indicated. 
Galaxies that are not detected in the $R$ band at a $3\sigma$ level
are plotted in the open circles by assigning them the $3\sigma$
limiting magnitude in $R$.  These indicate their lower-limit colours.
The solid line is the best-fit CMR, and the long dashed lines are offset
by $\Delta (R-z') = \pm 0.2$mag with respect to the best-fit line.
\label{fig:colmagRz}}  

\begin{center}
\leavevmode
\begin{center}
  \hspace*{-0.1cm}
  \epsfxsize 0.48\hsize
  \epsfbox{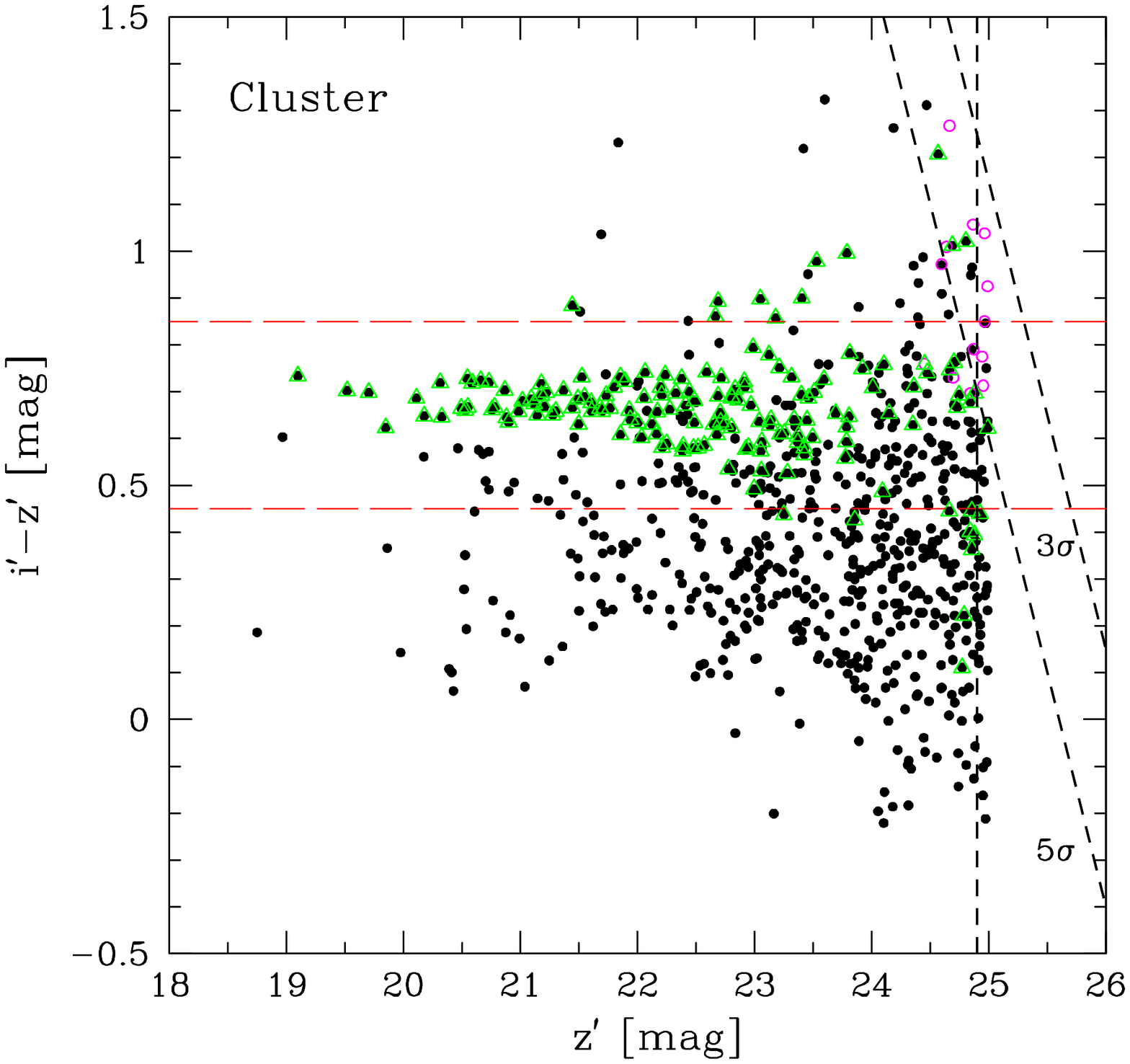}
  \epsfxsize 0.48\hsize
  \epsfbox{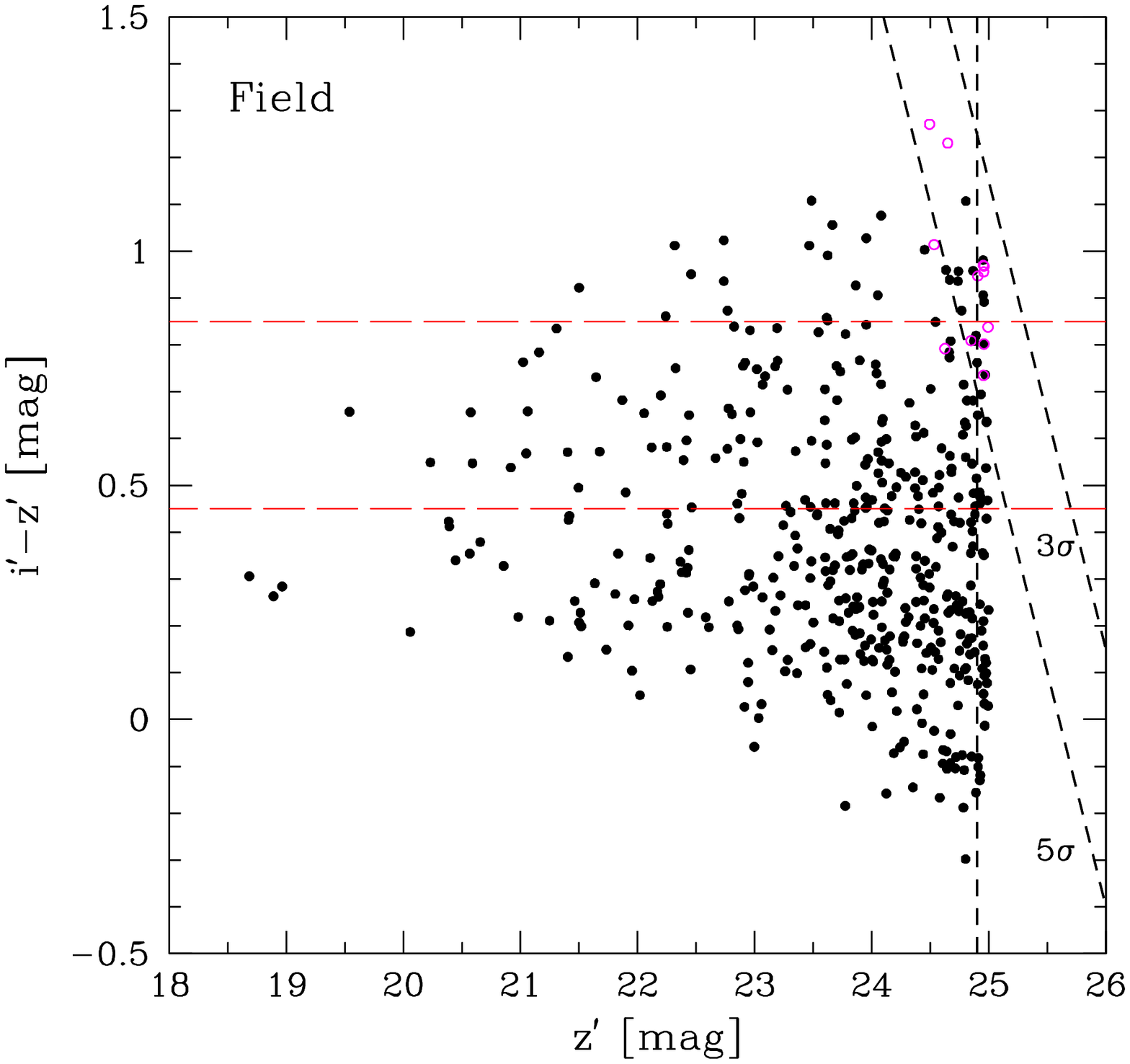}
\end{center}
\end{center}
\caption{Colour-magnitude diagrams ($i'-z'$ vs. $z'$) in the cluster 
({\it left }) and the field ({\it right}).  
The meanings of the lines are nearly
the same as in Fig.~\ref{fig:colmagRz}. In the left panel, galaxies 
on the red sequence in Fig. 5 are indicated as the triangles. The 
two horizontal dashed lines ($i' -z'=$ 0.45 and 0.85) show the 
colour selection criteria in the $i' - z'$ colour. 
\label{fig:colmagiz}}
\end{figure*}

In Figs.~\ref{fig:colmagRz} and \ref{fig:colmagiz},
we present the colour-magnitude diagrams of the ``cluster''
and the ``field'' regions defined in Section~\ref{subsec:largescale}. 
The field galaxies plotted in the right panels of
Figs.~\ref{fig:colmagRz} and \ref{fig:colmagiz}
are randomly sampled from the control fields by scaling down the
surface area to match the area of the cluster region.
We have to be very careful about the fact that some faint galaxies 
are detected at $\sim$5$\sigma$ level only in the $z'$-band, 
while not firmly detected in the other bands.
Since we cannot know the real colour of these faint galaxies, 
we plot them by open circles in Figs.~\ref{fig:colmagRz} and
\ref{fig:colmagiz} by assuming that they have the 
3$\sigma$ limiting magnitudes in the non-detected bands.
Therefore, the indicated colours of these galaxies are lower limits.

The CMRs are clearly seen in the cluster region in both colours, 
while they are virtually absent in the field region
(Figs.~\ref{fig:colmagRz} and ~\ref{fig:colmagiz}). 
We fit the $R-z'$ CMR, which is very sensitive to the 4000\AA\ break
at this redshift, as a linear relation of the form
$$ (R-z') = c_0 - \textrm{slope} \times  z' {} , $$ 
where $c_0$ is the zero point of the CMR.
We use only bright galaxies ($ z' < 22.5$) for the
fitting because photometric errors of bright galaxies are 
very small and they are not affected strongly by contaminant galaxies.
Also, we use only galaxies that have $R-z' > 1.0$ in fitting
the CMR of $R-z'$ vs. $z'$ since we should not include the blue
galaxies when fitting the red sequence. 
Based on the bi-weight fitting method,
we obtain the best-fitting CMR as shown by the solid line in
Fig.~\ref{fig:colmagRz}. This is expressed by:
\[
 (R-z') = (2.71 \pm 0.12) - (0.049 \pm 0.006) \times z' { }.
\]
This best-fitting CMR on the $R-z'$ vs. $z'$ diagram has a good agreement
with the visual impression. We also note that this is consistent with
the model predictions in \cite{kod97}.
Different method of fittings or different samples to use for fitting
may produce slightly different fitting results. However, we note  
that our conclusions are not affected by a small change of the slopes and
the zero points of the CMR. We also apply a colour selection in $i' -z'$
supplementarily to select the red galaxies. 
Because the slope of the $i'-z'$ CMR is consistent with zero,
we adopt a simple colour selection of $ 0.4 < i'-z' <0.8$.
The red galaxies located between the two dashed lines in Figs.~5 and 6 
are defined as ``red-sequence'' galaxies. We set these colour ranges 
broad enough (= 0.4 mag) so as not to miss the red member galaxies.
We apply a rather broad colour cut in $i'-z'$, 
since the 4000\AA\ break feature is neatly bracketed by the other colour,
$R-z'$, and we use the $i-z'$ colour only supplementarily.
In Fig. 6, galaxies on the red sequence defined in $R-z'$ colour in 
Fig. 5 are shown by triangles. We see that many of them are located in
the range of $0.45 < i' - z' < 0.85$, and our colour selection in $i'-z'$
rejects only a few galaxies which are located far from the red sequence in $i'-z'$.
However, at the very faint end, some red-sequence galaxies defined by $R-z'$
colours have relatively blue $i'-z'$ colours such as $<0.45$.
We will take care of them by applying even wider colour cut at the
faintest bin in the following discussion (see Section 3.3)

As will be discussed in detail in the following sections,
a deficit of the red-sequence galaxies is seen even by eye
at $m(z') \gsim 23.5$ ($\sim M^* +2 $ at the cluster redshift in the
case of passive evolution) in the cluster core, which is still
$\sim1.5$ mag brighter than the 5$\sigma$ limiting magnitude.
Similar deficits are reported in other $z \sim 0.8$ clusters
(\citealt{del07}; \citealt{tan05}; \citealt{del04}). 
In the next section, we quantify the deficit using luminosity functions.
  
\subsection{Luminosity Functions of the Red-Sequence Galaxies}
\label{subsec:LF}
\begin{figure}
   \begin{center}
    \leavevmode
    \rotatebox{0}{\includegraphics[width=9cm,height=9cm]{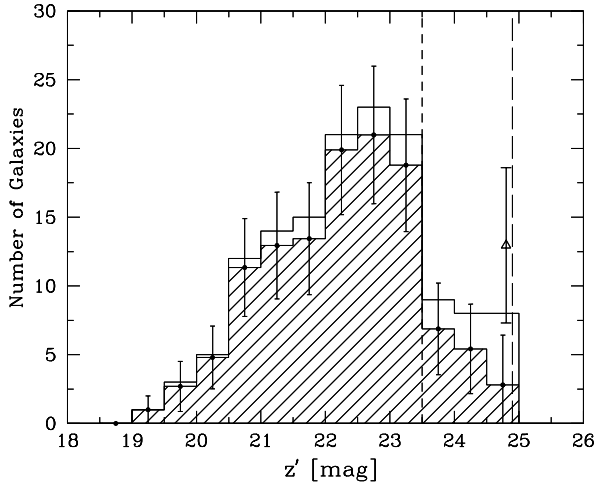}}
   \end{center}  
   \caption{Luminosity functions of the red-sequence galaxies in the
RXJ1716 cluster. The open
and shaded histograms show before and after the statistical field subtraction,
respectively.  Error bars represent the Poisson errors. The vertical
long-dashed line is the 5$\sigma$ detection limit 
same as Figs.\ref{fig:colmagRz} and \ref{fig:colmagiz}. The short-dotted 
line represents $m(z') = 23.5$, at which galaxies are classified 
into `luminous' and `faint'. The triangle at the faintest bin indicates 
the galaxy number count 
if we define the red sequence as $\pm 0.3$mag from the CMR.
\label{fig:LF}}
 \end{figure}

We present here the luminosity functions of the red-sequence galaxies in
the cluster region to quantify the deficit of faint red galaxies. 
In order to obtain the luminosity function of cluster ``member'' galaxies,
we should statistically subtract contaminant galaxies using the control
field sample.  Fig.~\ref{fig:LF} shows the luminosity functions of the
cluster region before ({\it open} histogram) and after 
({\it hatched} histogram) the statistical field subtraction.
Although the 5$\sigma$ limiting magnitude in the $z'$-band is
$24.9$ mag (shown as long dashed line in Fig.~\ref{fig:LF}), 
we plot luminosity functions down to $m(z') =25.0$.
We should also note that the faint objects 
plotted as open circles on the red sequence
in Figs.~\ref{fig:colmagRz} and \ref{fig:colmagiz} are all included
in these luminosity functions of the red-sequence galaxies to generously
correct for incompleteness. Because these galaxies may have redder colours
than shown, they can fall within the red-sequence cut.
Because of this, we could slightly overestimate the number of faint red
galaxies rather than underestimate it, which strengthens our conclusion of
the deficit of faint red galaxies.
It is very clear, even accounting for statistical errors, that there is a
deficit of faint red galaxies below $m(z') \sim 23.5 $. 
This clearly shows that the CMR is not fully in place in RXJ1716 at the
faint end.  This is qualitatively consistent with earlier results on
the RXJ0152 cluster (\citealt{tan05}) and on the EDisCS clusters
(\citealt{del04}, 2007). 

We also construct a luminosity function of the red sequence galaxies 
in the ``group'' environment, represented by the composite of the two
sub-clumps, and it is shown in Fig.\ 8. 
Although the statistics is poor and we cannot draw any firm conclusions,
it is interesting to note that a deficit of the faint red galaxies
can be even stronger and the number of faint red galaxies below $m(z')=23.5$
is consistent with zero after corrected for field contamination.

 \begin{figure}
   \begin{center}
    \leavevmode
    \rotatebox{0}{\includegraphics[width=9cm,height=9cm]{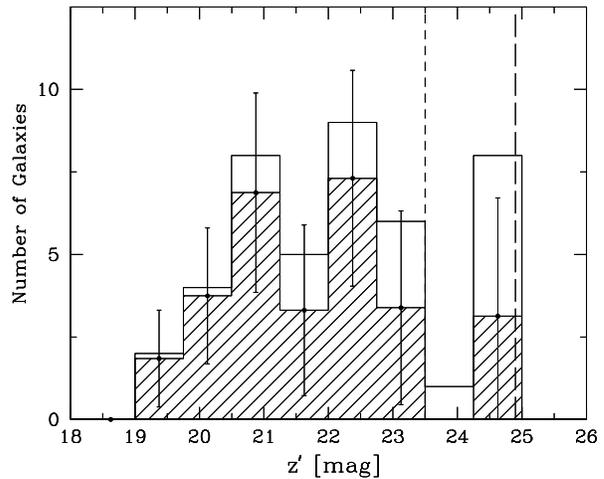}}
   \end{center} 
   \caption{Luminosity functions of the red-sequence galaxies in the
  ``group'' regions defined in Section 3.1. 
  The meanings of histograms and the lines are the
  same as in Fig.\ref{fig:LF}. Histograms are shown as a composite value of the 
  two groups. The bin size is 1.5 times wider than in Fig. 7 due to the
  small number of group galaxies.
\label{fig:LF_group}}
 \end{figure}
To quantify the ``deficit'' of faint red galaxies, we calculate 
the luminous-to-faint ratio of the red-sequence galaxies.
We separate luminous and faint galaxies at $m(z')=23.5$
(shown by the short-dashed line in Fig.~\ref{fig:LF}). 
We should note that since the typical photometric errors at the
faintest bin ($\sim 0.3$ mag) can be larger than the colour range
of the red-sequence galaxies ($\pm 0.2$~mag widths;
see Section \ref{subsec:CMD}), we may miss some red-sequence galaxies
at the faint end. Therefore, we also count the number of red-sequence
galaxies at the faintest bin ($24.5\le m(z') <25$) by applying
a wider colour range of the red-sequence galaxies, namely,
$\pm 0.3$ mag in both colours, which is shown by the triangle
in Fig.~\ref{fig:LF}.
Since the density of red galaxies is lower than that of blue
galaxies at the faint end, it is likely that the number of intrinsically
blue galaxies which would have entered the red-sequence from the
blue side due to photometric errors is larger than that of
intrinsically red galaxies which would have escaped from the
red-sequence if we apply such a broad colour cut.
As mentioned in Section. 3.2, because there are some $R-z'$ red-sequence
galaxies which have slightly bluer colour in $i'-z'$, 
the number counts at the faintest bin looks
jumped up when applying this wider colour cut.
However, the number of faint red galaxies should be taken as
an upper limit in this case, and hereafter we use this upper limit value
for the luminous-to-faint ratio so as not to underestimate it.
The luminous-to-faint ratio (lum/faint) is thus estimated to
be $4.3 \pm 1.4$.
In the following section, we compare this ratio to those of other
clusters at $z\sim0.8$ available in the literature.

\section{Discussion}
\label{sec:discuss}

For comparison, we summarize below some previous works on the faint end
of the CMR in $z\sim0.8$ clusters available in the literature.
We select only the surveys with a depth comparable to or deeper than
our study of the RXJ1716 cluster.
\begin{itemize}
\item {\bf RXJ0152}: This cluster at $z=0.83$ was studied in \cite{tan05}.
They defined the ``cluster'', ``group'' and ``field'' environments
based on the local and global densities, and
showed that the build-up of the CMR is delayed in lower density
environments. 
Their results suggest that the CMR appears earlier in cluster
environment, but the faint end of the CMR is not fully formed yet
at $z \sim 0.8$ even in the cluster region
when compared to the local SDSS data.
Therefore, a deficit of the faint red galaxies is seen
in this cluster.
Note that RXJ0152 consists of two major clumps (North and South). 
Both clumps are defined as ``cluster'' in Tanaka et al.\ (2005), 
and the luminosity functions shown in the paper are the
composite of the two clumps.
In their study, sources are detected in $z'$ band, and the limiting magnitude
is $m(z')=25.0$ ($\sim M^* + 3.5$). Rest-frame $U-V$ colour calculated 
using the redshifts of each galaxy is used 
to define the CMR, and the blue limit of the red sequence is set as 0.15 mag
bluer than the best-fit CMR. Cluster members are selected on the 
basis of photometric redshifts and then a statistical subtraction 
of the remaining contamination is made.
As a result, cluster member galaxies, which satisfy the conditions 
of the density, distribute within about virial radius from the centre
of the two clumps.  

\item {\bf MS1054}: This cluster at $z=0.83$ was 
studied in Andreon (2006). He showed that the faint end of the CMR 
is well visible in MS1054, and questioned the previous report
on the ``deficit'' of faint red galaxies at $z \sim 0.8$.
We should note that this cluster is very massive
as shown in Table~\ref{tab:cluster_data}. 
He performed a statistical subtraction of the field contamination
on the colour-magnitude diagrams, which is a similar method as we use for
RXJ1716. Goto et al. (2005) also studied this cluster.
They showed a deficit of faint red galaxies at 1$\sigma$ level 
based on spectroscopically confirmed member samples. 
However, their completeness is low at the faint end ($\sim$ 20 per
cent), and the statistical uncertainty is large.
Since the method of our analysis
is similar to that of Andreon (2006), we can compare our results
directly with Andreon's (2006). 
In his study, sources were detected in $I$ and $K$ band, 
and the limiting magnitudes reach $\sim M^* +3.5$ in both bands. 
The area studied as the cluster region is 4.3 Mpc$^2$. $V-I$ colour
is used to define the CMR, and the width of the red sequence 
is $\pm$0.3 mag from the best-fit CMR.

\item {\bf EDisCS clusters}:
Some clusters at $z=$ 0.7--0.8 from the ESO Distant Cluster Survey 
were studied in \cite{del04} and \cite{del07}. 
In both papers, a deficit of faint red galaxies is shown.  
However, we do not know the X-ray properties of these clusters
very well.  \cite{joh06} studied the X-ray properties of some 
of these clusters, but one of them turned out not to be a true
galaxy cluster. Judging from the X-ray data presented in \cite{joh06}
(CL1216-1201 and CL1054-1145), these clusters may be poorer systems
than RXJ1716, RXJ0152 and MS1054. The X-ray luminosities 
of CL1216-1201 and CL1054-1145 are
$ L_X \sim 5 \times 10^{44}$ erg s$^{-1}$ and 
$ \sim 2 \times 10^{44}$ erg s$^{-1}$, respectively.  
RXJ1716, RXJ0152 and MS1054 have larger $L_X$ (see Table~2).
In De Lucia et al. (2004, 2007), sources are detected in $I$ band, 
and the limiting
magnitude is $\sim M^* + 2.5$. $V-I$ colour is used to define the 
CMR, and the width of the red sequence is $\pm$ 0.3 mag from the
best-fit CMR. They calculate the $r_{200}$ for each cluster, and 
used the $\sim 0.5 \times r_{200}$ as the cluster radius.
They compared the results 
from the samples with and without the photometric 
redshift selection, and did not show much difference.
\end{itemize}  
 \begin{figure}
   \begin{center}
    \leavevmode
    \rotatebox{0}{\includegraphics[width=9cm,height=9cm]{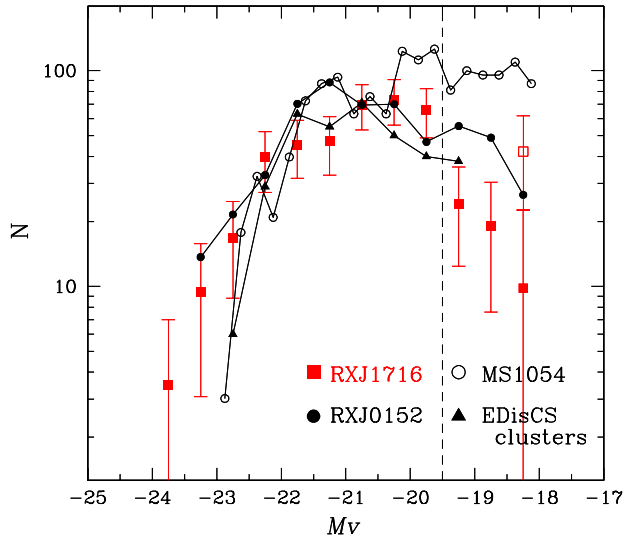}}
   \end{center} 
   \caption{Luminosity functions of galaxy clusters at $z\sim 0.8$.
The data are taken from Tanaka et al.~(2005) for RXJ0152, Andreon (2006)
for MS1054, and De Lucia et al.~(2007) for EDisCS clusters, respectively.
The meanings of the symbols are shown in the figure. The open square
indicates the number count of the faintest bin for RXJ1716 when
using the 0.6 mag width of the red sequences (see also Fig. 7).
The number of galaxies is normalized at $M_V = -20.75$.  
We present error-bars only for RXJ1716, and those for the other clusters
are omitted for clarity.
\label{fig:LF_hikaku}}
 \end{figure} 
\begin{table*}
\begin{center}
\begin{tabular}{c|c|c|c|c|c|c}
\hline
\hline
Cluster  & redshift &  $L_{\rm{bol}}$[$10^{44}$erg s$^{-1}$]  & 
$T_{\rm{gas}}$[keV] & lum-to-faint ratio & deficit  
\\
\hline
RXJ1716.4+6708 & 0.81 & $13.86 \pm 1.04$ & $6.8^{+1.0}_{-0.6}$ & $4.3 \pm
 1.4$ & Yes  
\\ 
RXJ0152.7-1357(S) & 0.83 & $7.73 \pm 0.40$ & $6.9^{2.9}_{-0.8}$  & $3.2 \pm
 0.5$ &  Yes   
\\
RXJ0152.7-1357(N) & 0.83 & $10.67 \pm 0.67$  & $6.0^{+1.1}_{-0.7}$  & ($3.2
 \pm 0.5$)& (Yes) 
\\ 
MS1054.4-0321 & 0.83 & $28.48 \pm 2.96$  & $10.2^{+1.0}_{-0.8}$  & $1.6
 \pm 0.2$ &  No 
\\

\hline
\end{tabular}
\caption{Summary of the properties of clusters.
The X-ray properties ($L_{\rm{bol}}$ and $T_{\rm{gas}}$)
are taken from Ettori et al.\ (2004).
Luminous-to-faint ratios are calculated by using the data from
Tanaka et al.\ (2005) for RXJ0152 and Andreon(2006) for MS1054. 
Note that RXJ0152 consists of two major clumps (N and S),
and the luminous-to-faint ratio is calculated for a composite
of the two.}
\label{tab:cluster_data}
\end{center}
\end{table*}

As summarized above, the definitions of cluster member galaxies and 
the colours used to define the CMR are not the same among different
studies of the $z \sim 0.8$ clusters.
However, as shown in \cite{del07}, the results would not be
significantly affected whether we do or do not apply photometric redshifts
for member selection. At the same time, the difference in the colours 
used to define the red sequence does not cause a problem when
selecting red galaxies at $z \sim 0.8$ because all colours 
used to define the red sequence in the papers have sensitivity 
to 4000\AA\ break feature at $z \sim 0.8$.
The studied surface area of the clusters differ from author to author.
To quantify this effect, we changed the radius of cluster regions for
RXJ1716 between $(0.35 - 1.0) \times r_{200}$, but we do not 
see any significant difference in the luminosity function within the error.
Therefore, the surface area of cluster region would not affect our 
conclusions strongly, 
as long as the dense region of each cluster is properly covered.
Below, we compare luminosity functions and luminous-to-faint ratios 
of the $z\sim0.8$ clusters.

In Fig.~\ref{fig:LF_hikaku}, we compare the luminosity functions of
the red-sequence galaxies of the above clusters available in the
literature. Note that the depths of the data for RXJ1716, RXJ0152 and MS1054
are about the same and $\sim 1$ mag deeper than that of the EDisCS clusters. 
We can see a decline of the luminosity function at $M_V > -20$ in RXJ1716,
RXJ0152, and the EDisCS clusters in \cite{del07}, while no such
decline is seen in MS1054. The ``deficit'' means that the faint end of
the CMR has not been fully established, while ``no deficit'' means the
faint end of the CMR has been already well populated by $z \sim 0.8$.
As claimed by \cite{and06}, the deficit of the faint red galaxies
may not be an universal phenomenon.

We calculate the luminous-to-faint ratios of the red-sequence galaxies
for RXJ0152 and MS1054 using the same definition of ``luminous'' and
``faint'' galaxies as defined in Section~3.3. 
The dividing magnitude of ``luminous'' and ``faint'' is $m(z') =23.5$
and it corresponds to $M_V \simeq -19.5$.
We obtain lum/faint= $3.2 \pm 0.5$ for RXJ0152 and 
$1.6 \pm 0.2$ for MS1054.
The numbers of luminous and faint galaxies of RXJ0152 
are taken from \cite{tan05}, and those of MS1054 are read off 
from the number counts plotted in Fig.~4 of \cite{and06}.
Note that the luminous-to-faint ratio
for EDisCS clusters in \cite{del07} is not shown here because their
data are not deep enough for us to calculate the luminous-to-faint
ratio in the same definition
(our data is $\sim$1 mag deeper than those in \citealt{del07}). 
In terms of luminous-to-faint ratio, RXJ0152 has a similar value to 
that of RXJ1716 ($4.3 \pm 1.4$)
within errors, while MS1054 has a significantly smaller value.
We also obtain lum/faint $\simeq 1.0 \pm 0.1$ for the Coma cluster
in the same definition,
using the luminosity function of red-sequence galaxies of the 
Coma cluster shown in Fig.~7 of \cite{del07}.
Therefore, the local rich cluster Coma has even smaller luminous-to-faint
ratio than those of any $z \sim 0.8$ clusters discussed in this paper including
the richest cluster MS1054 (see below and Table~2).
In fact, Coma has a "rising" faint end in the luminosity function of 
red galaxies, compared to the flat or declining faint ends seen 
in MS1054 and the other $z \sim 0.8$ clusters (Fig.\ 9).
This may indicate evolution in the faint end of luminosity functions 
of red galaxies between $z \sim 0.8$ and the present day. However, 
we do not go further on this issue in this paper, because comparing 
clusters at different redshifts is not trivial, given the expected 
evolution in the richness of clusters themselves by accretion 
of surrounding systems.

\begin{figure}
   \begin{center}
    \leavevmode
    \rotatebox{0}{\includegraphics[width=9cm,height=9cm]{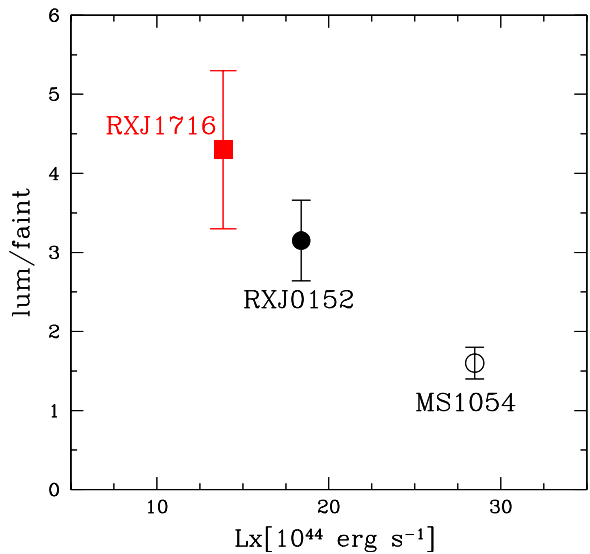}}
   \end{center} 
   \caption{Luminous-to-faint ratio of $z \sim 0.8$ clusters against
their X-ray luminosity. The error-bars
represent the Poisson error. Note that the $L_X$ for RXJ0152 is shown as a
composite of the two clumps.
   \label{fig:lum/faint}}
 \end{figure}

Here, we focus on the difference in the faint end of MS1054 from
the other clusters at $z \sim 0.8$.  What causes this difference?
MS1054 is a very rich, massive cluster, and
this may have some impact on the build-up of the CMR.
Cluster richness and mass are well correlated with the
X-ray properties such as luminosity $L_X$ and temperature $kT$. 
\cite{ett04} presented the X-ray properties of RXJ1716, RXJ0152 and
MS1054 clusters based on Chandra data.
We summarize them in Table~2. In Fig.~\ref{fig:lum/faint},
the luminous-to-faint ratios are plotted against the X-ray luminosities.
Although only three clusters are currently available, 
it is interesting to note
that the richest system with the highest $L_X$, MS1054, has the lowest 
luminous-to-faint ratio.

A similar trend was seen in a super-rich cluster at a slightly lower redshift,
the CL0016 cluster at $z=0.55$ \citep{tan05}.
In fact, this cluster is one of the richest systems ever known at intermediate
redshifts and it has $L_X =53.27\pm 7.33 \times 10^{44}$~erg s$^{-1}$
and $kT = 10.0 \pm 0.5$~keV (\citealt{ett04}). 
Interestingly, this cluster shows no deficit of the faint red 
galaxies in the cluster core, which is the same trend as the 
rich cluster MS1054.
The fact that we see no deficit in CL0016 may not be simply 
due to its lower redshift or later evolutionary stage,
but it is also possible that rich systems tend to have many 
faint red galaxies. 

We therefore suggest that the build-up of the CMR is 
dependent not only on redshift
but also on cluster richness, in the sense that richer systems have had
earlier galaxy evolution even for faint galaxies and the faint end of
the CMR
has already been well developed.  This is similar to what Tanaka et
al.\ (2005)
found in their analysis by separating galaxies into three environmental
bins (cluster/group/field). 
In the current analysis, we now suggest that even within the cluster
environment, the evolutionary stage of faint galaxies still depends on
the richness of clusters.

\cite{del07} recently investigated the environmental dependence
of the luminous-to-faint ratios of the red sequence galaxies by
dividing the clusters into two classes according to their velocity
dispersions at $600$ km s$^{-1}$.
However, the luminous-to-faint ratios of the two classes do not
show a significant difference.
It should be pointed out, however, that the RXJ1716, RXJ0152 and MS1054
clusters studied in this paper have much larger velocity
dispersions compared to the EDisCS clusters.
The velocity dispersion is $\sim 1500 $ km s$^{-1}$ for RXJ1716 (see
Section 2.3), $\sim 1200$ km s$^{-1}$ for MS1054 (Tran et al. 1999), 
$\sim 900$ km s$^{-1}$ for RXJ0152(N) and  
$\sim 700$ km s$^{-1}$ for RXJ0152(S)
(Demarco et al. 2005). The velocity dispersion of RXJ0152 estimated from 
all the cluster galaxies is $\sim 1600$ km s$^{-1}$ (Demarco et al.~2005).

Difference in the faint end of the red sequence may be seen only
at the richest end of clusters, and the CMR may be fully established
toward the faint end only in the super rich environment at $z \sim 0.8$. 
In order to confirm the relationship between the ``deficit'' of faint
red galaxies and cluster richness, 
we definitely need a larger sample of clusters at high redshifts.
\section{Summary and Conclusions}
\label{sec:summary}
Using a deep, multi-colour, panoramic imaging data set of the distant 
cluster RXJ1716.4+6708 at $z=0.81$, newly taken with the Prime Focus 
Camera (Suprime-Cam) on the Subaru Telescope, we have carried out an analysis 
of red-sequence galaxies
with a care for incompleteness. We have found that there is 
a sharp decline in the number of the red galaxies toward 
the faint end of the CMR below $M^*+2$.
We compared our results with those for other clusters at $z \sim 0.8$ 
taken from the literature,
by calculating the luminous-to-faint ratio to quantify the degree of
the ``deficit'' and by combining the information on richness of the 
individual clusters from X-ray properties.  We suggest that
the deficit of faint red galaxies is dependent on the richness
or mass of the clusters in the sense that poorer systems show stronger
deficits.  This indicates that the evolutionary 
stage of less massive galaxies depends critically on environment.
In order to confirm this interesting trend, we need a much larger sample of
galaxy clusters over a wide range in richness, and not only at similar
redshifts but also at other redshifts. 

\section*{Acknowledgment}

We thank the anonymous referee for the careful reading of the paper and
for helpful suggestions, which improved the paper.
This work was financially supported in part by a Grant-in-Aid for the
Scientific Research (Nos.\, 15740126; 18684004) by the Japanese 
Ministry of Education, Culture, Sports and Science. 
This study is based on data collected at the Subaru Telescope, which 
is operated by the National Astronomical Observatory of Japan. 
M.T. acknowledges support from the Japan Society for the Promotion 
of Science (JSPS) through JSPS research fellowships for Young Scientists.


\end{document}